\documentclass[12pt]{iopart}

\usepackage{graphicx}

\begin{document}

\title[Resonant photoemission and electronic structure of 1T-Mn$_{0.2}$TiSe$_2$]{Resonant photoemission at the L$_3$ absorption edge of Mn and Ti and electronic structure of 1T-Mn$_{0.2}$TiSe$_2$}

\author{M.V. Yablonskikh$^1$, A.S. Shkvarin $^2$, Y.M. Yarmoshenko$^2$, N.A. Skorikov$^2$, A.N. Titov$^2,^3$}

\ead{mikhail.yablonskikh@gmail.com}

\address{$^1$Sincrotrone Trieste SCpA, Basovizza I-34012, Italy} 
\address{$^2$ Institute of Metal Physics, Russian Academy of Sciences-Ural Division, 620041 Yekaterinburg, Russia}
\address{$^3$Ural Federal University, pr. Lenina 51, Yekaterinburg, 620089 Russia }

%
\begin{abstract}
Resonant valence-band X-ray photoelectron spectra (ResPES) excited near 2p$_{3/2}$ core level energies, 2p X-ray photoelectron spectra (XPS) and L$_{3,2}$ X-ray absorption spectra (XAS) of Ti and Mn in single crystal of 1T-Mn$_2$TiSe$_2$  were studied for the first time. The ionic-covalent character of bonds formed by Mn atoms with the neighboring Se atoms in the octahedral coordination is established. From the XPS and XAS measurements compared with results of atomic multiplet calculations of Ti and Mn L$_{3,2}$ XAS  it is found that Ti  atoms are in ionic state of  4+ and Mn atoms are in the state of  2+.  In ResPES of Mn$_{0.2}$TiSe$_2$ excited near Ti  2p$_{3/2}$ and Mn 2p$_{3/2}$ absorption edges the Ti 3d and Mn 3d bands at binding energies just below the Fermi level are observed. According to band structure calculations E$(\textbf{k})$  the Ti 3d states are localized in the vicinity of $\mathsf{\Gamma}$ point and the Mn 3d states are localized along the direction K-$\mathsf{\Gamma}$-M  in the Brillouin zone of the crystal. 
 
 \end{abstract}

\pacs{ 74.25.Jb, Electronic structure; 
 31.10.+z,Theory of electronic structure, electronic transitions, and chemical bonding;
71.20.Be, Transition metals and alloys;
 78.70.Dm, X-ray absorption spectra;
 79.60.-i, Photoemission and photoelectron spectra}

\submitto{\JPCM}
\maketitle

\section{Introduction}
\label{intro}

Layered dichalcogenides of Ti intercalated with transition metals TM$_x$TiZ$_2$ (TM-transition metal, Z = S, Se, Te) are promising materials for the creation of monoatomic layers of magnetic metals with a given density \cite{Friend:1987p14556},\cite{Inoue:1989p14699} where these layers are separated by nonmagnetic slabs of the TiZ$_2$ matrix. Although many basic physical and chemical properties of TiZ$_2$ dichalcogenides are defined by the chemical type of the chalcogenide element Z and it's interaction with Ti \cite{Bullett:1978p6737}, the most interesting changes in physics and chemistry take place in TM$_x$TiZ$_2$ compounds. TM$_x$TiZ$_2$ intercalates are of fundamental scientific interest because of formation of magnetic phase transitions \cite{Tazuke:2006p14463}  and formation of charge density waves (CDW)  \cite{Rossnagel:2011p14467}.  Discovery of CDW to superconductive state transition \cite{Morosan:2006p5364} in TiSe$_2$ intercalated with copper and expectation of such in TiSe$_2$ intercalated with iron and chromium \cite{Titov:2006p5343} along with emerging progress in synthesis of single crystals of TM$_x$TiZ$_2$ and expected dependencies of their physical properties both from the type of intercalated TM and it's concentration $x$ have inspired a growing number of studies of electronic structure of those compounds \cite{Jeong:2010p14519}.  

It was established that electronic properties of TM$_x$TiZ$_2$ can be understood in terms of charge transfer and interlayer separation \cite{Friend:1987p14556}. In the charge transfer process a  strong hybridization between 3d orbitals of intercalated TM atom and Ti 3d along with Z p orbitals plays a major role resulting in relative energy shifts of element-specific electronic bands and changes of their localization degree, energy shift of conduction band and variation of density of states at the Fermi level E$_F$ \cite{Inoue:1989p14699}.  Since the interlayer separation is dependent from the magnitude of \textit{c$_0$} lattice constant,  it was suggested that the increase of spacing between Se-Ti-Se tri-layers is a result of influence of the hybridization of TM 3d-electrons with Ti 3d and Se 4p electrons, where the strength of the hybridization degree is dependent from the type and concentration of intercalated TM  \cite{Maksimov:2004p3818}.  From a number of studies about chemical bonding in various TM$_x$TiZ$_2$ compounds it is already known that hybridization of 3d states of an intercalated TM atom and its nearest neighborhood is to evidence the covalent origin \cite{Titov:2001p14294,Postnikov:2000p3727,Titov:1998p6002}  of chemical bonding in the majority of intercalated materials. In brief, the covalence of chemical bonding in TM$_x$TiZ$_2$ results in the compression of the lattice in the \textit{c} direction shifting the tri-layers closer to each other along a normal line to a basal plane  \cite{Titov:2000p3790}, decrease of the conductivity  \cite{Titov:2001p14294}, formation of narrow and dispersionless bands in the vicinity of the Fermi level \cite{Kuznetsova:2005p5999,Cui:2006p5774} and a shift of the Fermi level \cite{Titov:2010p8966} relatively to the energy position of the original TiSe$_2$ band. Suppression of magnetic moments of intercalated TM atoms \cite{Titov:2004p3816} is due to the variation of the localization degree of TM 3d, Ti 3d electrons and the hybridization of those with Z element 4p states.  

The only exception in the whole family of TM$_x$TiZ$_2$ is the Mn$_x$TiSe$_2$ branch which exhibits both growth of conductivity \cite{Maksimov:2004p3818} and anomalously high increment of the Pauli contribution to the magnetic susceptibility  when compared with other TM$_x$TiZ$_2$ \cite{Tazuke:2006p14463}. First, the magnitude of lattice constant \textit{c$_0$} is the largest amongst both TiSe$_2$\cite{Maksimov:2004p3818} and TiSe$_2$ intercalated with Co or Cr \cite{Titov:2001p14294}. It is also to be noted that the \textit{c$_0$} in TiSe$_2$ is smaller than that in Mn$_x$TiSe$_2$ but larger than that both in Co$_x$TiSe$_2$ and Cr$_x$TiSe$_2$. In other words,  the interlayer spacing between the tri-layers of TiSe$_2$ in Mn$_x$TiSe$_2$ is increased in \textit{c} direction as a result of intercalation of Mn between the tri-layers.  Second, the hybridization between Mn and TiSe$_2$ lattice is assumed to be of the lowest magnitude in the whole line of TM$_x$TiZ$_2$  \cite{Maksimov:2004p3818}. Third, effective magnetic moment  of Mn in Mn$_x$TiSe$_2$ is found to be of highest magnitude \cite{Maksimov:2004p3818},\cite{Postnikov:2000p3727} then those of other TM in TM$_x$TiZ$_2$. Those interesting results are reported for Mn$_x$TiSe$_2$ compounds where Mn concentration $\textit{x} \leq 0.2$. It is also expected that the bonding situation in Mn$_x$TiSe$_2$ will be quite outstanding from that in non-Mn TM$_x$TiSe$_2$ compounds \cite{Shkvarin2011-2}. 

Here we present experimental and theoretical research of electronic structure of Mn$_{0.2}$TiSe$_2$ single crystal. The primary goal is to obtain an element specific information about chemical bonding in Mn$_x$TiSe$_2$. The secondary goal is  to evaluate a localization degree of Mn 3d and Ti 3d electrons. The third goal is to reveal the details of charge transfer mechanism between intercalated atoms of Mn and those of TiSe$_2$ lattice.  Ti 2p and Mn 2p core level X-ray photoelectron spectra (XPS), Ti L$_{3,2}$ and Mn L$_{3,2}$ X-ray absorption spectra (XAS) along with valence band resonant photoelectron spectra (ResPES) excited across L$_3$ edges  of Ti and Mn have been measured. Experimental results are compared with spectral simulations, density of states  (DOS) and E$(\textbf{k})$ calculations.
 
\section{Experiment and calculations}
\label{sec:Experiment and calculations}

\begin{figure*}[htbp]
\begin{center}
\includegraphics[width=0.8\textwidth]{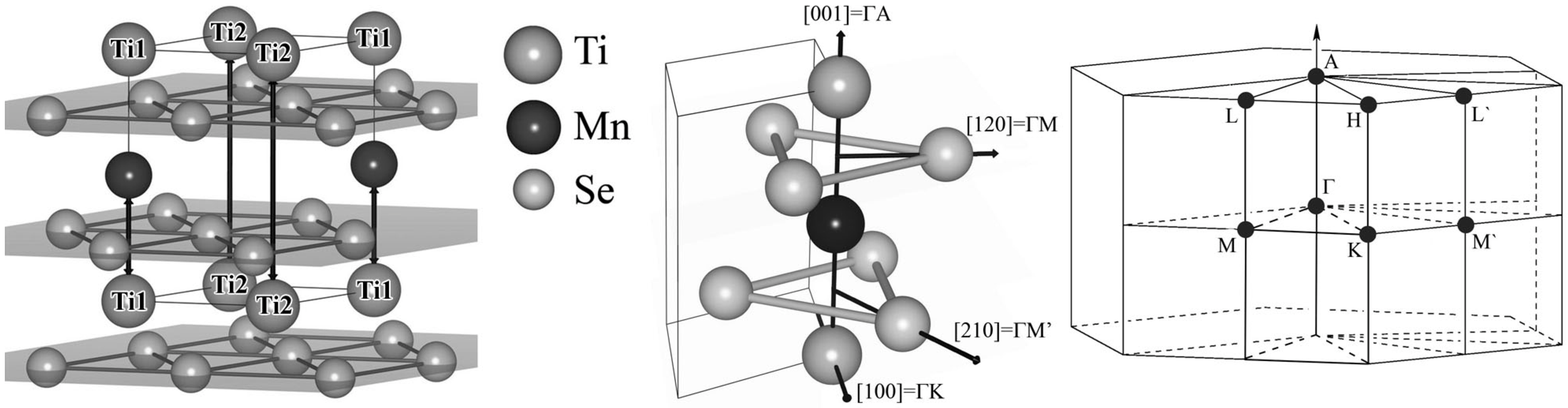}
\end{center}
\caption{A fragment of the Mn$_{0.2}$TiSe$_2$ crystal and Brillouin zone for the 1T-TiSe$_2$ compound. The $\mathsf{\Gamma}$-A direction in the Brillouine zone corresponds to the direction of the crystallographic \textit{c} axis perpendicular to the basic plane of the crystal in the direct space ([001] direction.}
\label{fig:1}       
\end{figure*}

Polycrystalline samples were synthesized by the ampoule synthesis technique. Powder diffractometry together with magnetic measurements showed  good sample quality in accordance with available literature data \cite{Titov:2004p3816,Maksimov:2004p3818,TAZUKE:1997p6390}. Single crystals of Mn$_x$TiSe$_2$ were grown by the gas-transport reaction technique with use of I$_2$ as a gas-carrier in evacuated quartz ampoules from a polycrystalline phase. Crystals have a form of thin plates of about $2 \times 1 \times 0.05$ mm$^3$ in size. The chemical composition was determined by X-ray fluorescence analysis using a JEOL-733 spectrometer. The crystal structure is the same as  reported earlier \cite{Titov:2004p3816}. Analysis showed the preservation of composition of the initial phase during the growth of the single crystals. A fragment of the crystal structure and the Brillouin zone of the compound 1T-Mn$_x$TiSe$_2$ is shown in Fig. \ref{fig:1}.


Spectroscopic measurements were performed at BACH beamline \cite{Zangrando:2001p3860} of Elettra synchrotron facility. X-ray photoelectron spectra (XPS) were taken using an VSW 150 photoelectron analyzer with an acceptance angle of 8 degrees. X-ray absorption data were measured in the Total Electron Yield (TEY) mode. Samples were cleaved in-situ in a vacuum 3-5  $\times$ 10$^{-10}$ Torr. XPS of the C 1s, O  1s and Ti 2p core levels were measured from time to time to check for a possible contamination of the surface during the measurements. Intensity of carbon and oxygen 1s peaks in survey spectra was extremely low and the shape of Ti 2p XPS remained unchanged during the measurement. Absence of titanium oxide like satellite peaks in Ti 2p XPS ensures the negligible influence of probable oxidation both of the sample surface and of the bulk material. The angle between the incoming beam and the normal to the basis plane of the sample was 60 degrees. The normal to sample coincides with the \textit{c} axis [001]  which was aligned with the axis of the analyzer. Both the \textit{c} axis of the sample and the polarization of the incident X-ray were in horizontal plane. The energy resolution of the monochromator i.e. the energy resolution of X-ray absorption spectra of Ti 2p and Mn 2p was set to 0.1 eV. Photoelectron analyzer resolution was set to 0.147 eV resulting in a total energy resolution of 0.19 eV both for XPS and resonant ResPES. Binding and monochromator energy scales were calibrated through the binding energy of Au 4f$_{7/2}$ level and through the Fermi edge measurements.

Band structure calculations were performed using the full-potential augment plane-wave method as implemented in WIEN2k code \cite{Blaha}. To account for the exchange-correlation potential we employed the gradient approximation \cite{Perdew:2010p6391} in the Perdew-Burke-Ernzerhof variant. The Brillouin zone integrations were performed with a $7\times 7\times\ 3$  special $\textbf{k}$- point grid and R$_{MT}^{min}\cdot$K$_{max}$=7 (the product of the smallest atomic sphere radii R$_{MT}$ and the plane wave cutoff parameter K$_{max}$) was used for the expansion of the basis set. The experimentally determined lattice parameters of  Mn$_{0.145}$TiSe$_2$ ($a_0$=3.5770\AA, $c_0$=6.1645 \AA) were taken. The spheres radii were chosen as $R_{Mn}$=2.46, $R_{Ti}$=2.5, $R_{Se}$=2.19 a.u. in a way that the corresponding spheres were near the mutual touch. To account for an intercalation of TiSe$_2$ by Mn atoms we constructed a $2a_0 \times 2a_0 \times 2c_0$ supercell in which for one from non-equivalent Se atoms the Z parameter was chosen as in pristine TiSe$_2$ whereas for another Se atoms (situated near the Mn layer) the experimentally determined Z-parameter was used. The resulting unit cell has the same symmetry as the initial unit cell of the TiSe$_2$: space group P-3m1 (164), see also Fig. \ref{fig:1}. According to the results of the calculations Mn 3d electrons are spin-polarized and the magnetic moment at Mn atoms reaches 3 $\mu_B$, that is found to be in a good agreement with previously reported experimental results  \cite{Maksimov:2004p3818} and calculations \cite{Postnikov:2000p3727}.

Ti atoms in TM$_x$TiZ$_2$ compounds can be classified in two types  further labelled as Ti1 and Ti2.  The nearest neighborhood of Ti2 atoms is identical to that in the matrix of TiSe$_2$ while for Ti1 atoms the neighborhood is changed. Atoms of intercalated Mn randomly occupy some sites in the second coordination shell of Ti1 as shown in Fig. \ref{fig:1}. The point group symmetry of titanium atoms in TiSe$_2$ is D$_{3d}$. The crystal field splits Ti 3d level into A$_{1g}$($3z^2-r^2$), E$_g^\pi$ ($x^2-y^2$, $xy$) and E$_g^\sigma$ ($xz$, $yz$) sublevels. As a result of intercalation the symmetry of Ti atoms is lowered to the C$_{3v}$ point group for Ti1 atoms and C$_s$ for Ti2. Taking into account that the nearest neighborhood of Ti (chalcogenic octahedra) is not changed during intercalation and the defects, which lower the symmetry, occur both in a second coordination sphere of Ti2 atoms and in further spheres of Ti1 atoms, the calculation results are given in terms of the D$_{3d}$ group. Our cluster calculation are restricted to use distorted octahedral coordination of investigated atom. For a comparison of results of cluster and band structure calculations please note two following remarks.
The analogue of cubic t$_{2g}$ states in octahedral coordination are A$_{1g}$ and E$_g^\pi$ states of trigonal basis (lobes of corresponding orbitals do not point towards atoms of nearest neighborhood). The analogue of cubic E$_g$ are E$_g^\sigma$ states of trigonal basis (lobes of corresponding orbitals point towards atoms of nearest neighborhood).

Atomic multiplet calculations of L$_{3,2}$ XAS of Ti$^{4+}$  and Mn$^{2+}$ in the octahedral coordination were performed using the program written by F. de Groot \cite{Stavitski:2010p8623}. For the simulated Ti L$_{3,2}$ absorption spectrum the best agreement with the experiment is achieved by the following parameters: Slater integrals are decreased to 75\% of the calculated value, the parameter of splitting in crystal field 10Dq is set to 1.2 eV, the charge transfer parameter $\Delta$ is set to 3 eV. For the simulated Mn L$_{3,2}$ absorption spectrum the crystal field splitting 10Dq is set to 0 eV, Slater integrals are set to 100\% of the calculated value, the charge transfer parameter $\Delta$ is set to 3 eV. The same values of the parameters were used to calculate Mn 2p core level photoelectron spectrum. Since Mn$_x$TiSe$_2$  compounds with concentrations  $x\leq0.25$  can be considered as two-dimensional \cite{Maksimov:2004p3818},  our experimental results can be discussed within an assumption of two-dimensionality taking into account that without ordering in Mn sub-lattice, crystal structure of these compounds is considered as a hexagonal.

\section{Experiment, Theory and Discussion}
\label{sec:Discussion}

\subsection{XPS}
\begin{figure*}[htbp]
\begin{center} 
\includegraphics[width=0.75\textwidth]{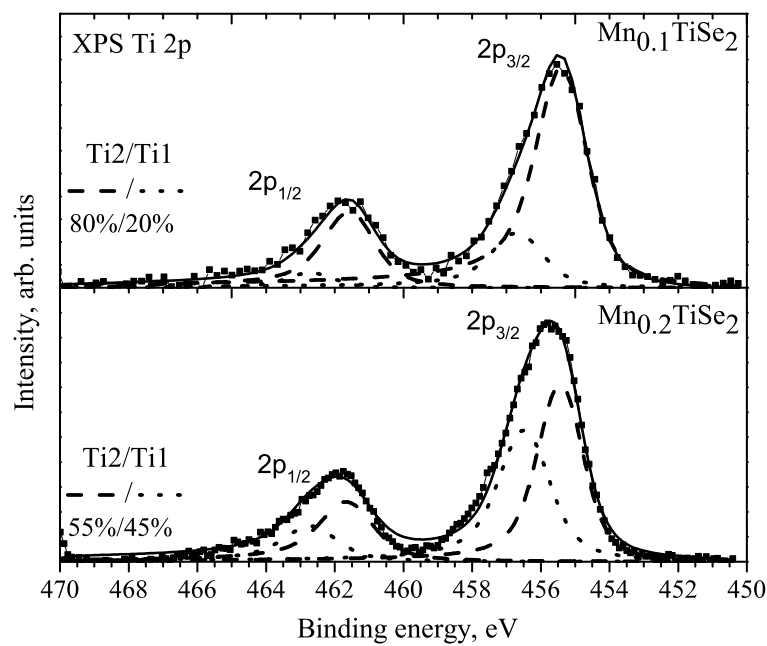}
\end{center} 
 \caption{XPS of Ti 2p  core levels in Mn$_x$TiSe$_2$ ($x$=0.1 and 0.2) and their fits are presented. Spectral contributions of Ti2 and TI1 atoms are shown by a long dash line ( \broken) and short dash line (\dashed) respectively at both top and bottom frames of the graph. Corresponding Ti2/Ti1 ratio numbers are given.}
\label{fig:2}       
\end{figure*}

 The results of measurements of Ti $2p$ XPS are shown in Fig. \ref{fig:2}. The binding energy of Ti 2p$_{3/2}$ maximum for Mn$_{0.2}$TiSe$_2$ is about 455.5 eV for x=0.1 and 455.8 eV for x=0.2 while for TiSe$_2$ it is equal to 455.0 eV \cite{Titov:2001p14294}.  Ti 2p core level peaks in the spectral multiplet of the intercalated compound are essentially broadened in comparison with that for pure TiSe$_2$  \cite{Titov:2001p14294},\cite{AShkvarin}. This is  a result of formation of two types of Ti atoms which occupy correspondingly two nonequivalent types of crystallographically different positions described in details in the Sec.  \ref{sec:Experiment and calculations}. Fits of the experimental Ti 2p XPS of Mn$_{x}$TiSe$_2$ for concentrations $x$=0.2 and 0.1 were performed under the assumption that both theoretical and experimental spectra consist of two spectral contributions from Ti1 and Ti2 sites and the spectral contribution from Ti2 atoms will prevail over that from Ti1 atoms. Although fitted XPS shown in Fig. \ref{fig:2} can not precisely reproduce the experimental spectra, a principal agreement between experimental XPS and the fits is achieved. As can be easily seen there is a correlation between the weighted components of the fits of Ti 2p XPS  and the concentration $x$ of intercalated Mn. Initial weight ratio of Ti2/Ti1 contributions  in Ti 2p XPS  is 80/20 for $x$=0.1. It changes approximately to 55/45 for $x$=0.2, i.e. approximately of factor two which corresponds to the doubling of Mn concentration. For these reasons the appearance of the contribution from Ti1 atoms in Ti 2p XPS is a result of intercalation. The energy difference of the magnitude about 1 eV between the Ti1 and Ti2 spectral weights maxima indicates that intercalation of Mn, even for cases of small concentrations, greatly affects electronic structure of TiSe$_2$ host lattice. 

\begin{figure*}
\begin{center} 
\includegraphics[width=0.75\textwidth]{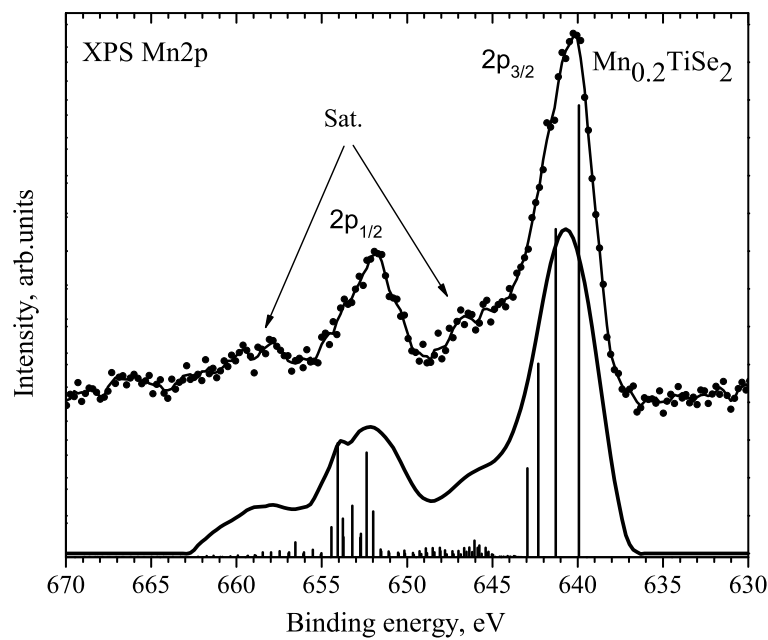}
\end{center}
\caption{Mn 2p X-ray photoelectron spectrum of Mn$_{0.2}$TiSe$_2$ is shown at the top of the figure. Positions of the satellites are marked with the label "Sat." Calculated spectrum is shown at the bottom part of the figure.}
\label{fig:3}       

\end{figure*}

 Core level 2p XPS of TM in TM$_x$TiZ$_2$ are most informative about physical properties and chemical bonding \cite{Titov:2001p14294}. The experimental Mn 2p photoelectron spectrum shown in Fig. \ref{fig:3} exhibits satellite structures at the binding energy about 6 eV higher than the peak energies of Mn 2p$_{3/2}$ and Mn 2p$_{1/2}$ core levels. Multiplet calculation of Mn 2p photoelectron spectrum of Mn$^{2+}$ shown in Fig. \ref{fig:3} also exhibits the presence of satellites at the distance of about 6 eV higher from the energy of theoretically computed Mn 2p$_{3/2}$ and 2p$_{1/2}$ peaks.  According to the results of the simulation both 2p and satellite peaks are formed mostly by mixing final  states 2p$^5$3d$^6$$\underline{L}$  and 2p$^5$3d$^5$ correspondingly. The shape of Mn 2p photoelectron spectrum is very close to that observed for Mn$_{1/4}$TiS$_2$ \cite{PhysRevB.46.3771}.  Calculation of Mn 2p X-ray photoelectron spectra performed for Mn$_{1/4}$TiS$_2$ also demonstrates the presence of similar satellite structure \cite{PhysRevB.38.3676}. This match between experimental and computational data of Mn intercalated diselenide and disulphide indicates that a charge transfer occurs between manganese atoms and ligands.  According to calculations  \cite{PhysRevB.38.3676}  for disulphide the charge-transfer energy $\Delta = 4$ while in presented calculations of diselenide $\Delta=3$. That difference in the magnitude of the  charge transfer energy means that the ionic component of chemical bonding in disulphides is larger than that in diselenides \cite{AShkvarin}. The fact that the satellite peaks were not observed in 2p XPS of Cr and  Co intercalated in TiSe$_2$ \cite{Titov:2001p14294,Postnikov:2000p3727} indicates a negligible charge transfer in Co$_x$TiSe$_2$ and Cr$_x$TiSe$_2$ compounds and proves that in Mn$_{0.2}$TiSe$_2$ the chemical bonding of Mn with the neighborhood is partially ionic.

\subsection{XAS}
\label{sec:XAS}
\begin{figure*}
\begin{center} 
\includegraphics[width=0.98\textwidth]{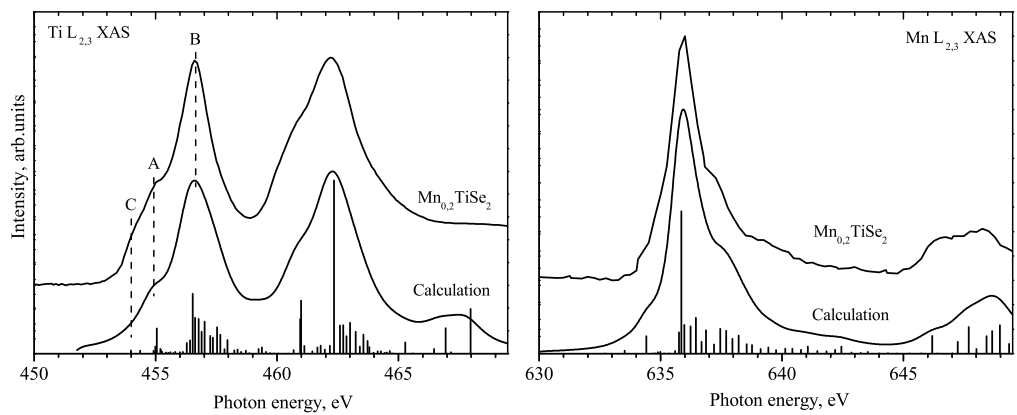}
\end{center}
\caption{Experimental (top) and calculated (bottom) Ti L$_{3,2}$ and Mn L$_{3,2}$ X-ray absorption spectra of Mn$_{0.2}$TiSe$_2$ are shown in the left and the right side of the figure.}
\label{fig:4}       
\end{figure*}

Let us discuss Ti L$_{3,2}$ and Mn L$_{3,2}$ XAS (Fig. \ref{fig:4}) measured in the same experimental conditions as XPS. The experimental Ti L$_{3,2}$ X-ray absorption spectrum is different from that of pristine TiSe$_2$ which was measured earlier \cite{Titov:2004p5361}. The difference is a result of the influence of intercalated Mn on the local neighborhood. Intercalation of Mn atoms in the Van der Waals gap of the TiSe$_2$ matrix, which results in the deformation of chalcogenic surrounding of Ti atoms and the charge transfer from Mn into the conduction band, is the main reason for differences of  both the peak shape and  the energy shift of  Ti 2p absorption spectrum from those of TiSe$_2$ \cite{Titova:2008p8695}. In Mn$_x$TiSe$_2$ where $x\leq0.2$ an increase of the magnitude of the lattice constant along the \textit{c} direction estimated relatively to that of TiSe$_2$ takes place in the neighborhood of Ti1 atoms. The amplitude of the deformation caused by this distortion of chalcogen octahedra is described by the parameter $\varepsilon=(2zc_0)/a_0$, where $c_0$ and $a_0$ are lattice constants, and Z is a coordinate of Se atom in the unit cell. The intercalation results in a change of the $\varepsilon$ value from 0.8655 in TiSe$_2$ to 0.8792 ($\Delta\varepsilon$= 0.0137) for Mn$_{0.2}$TiSe$_2$. It has to be noted that the amplitude of the deformation $\Delta\varepsilon$ has a quite large magnitude for case of Mn$_x$TiSe$_2$ in comparison with the cases of intercalation of other metals. For example, intercalation of Cr results in the quantity $\varepsilon$= 0.8651 ($\Delta\varepsilon$=0.0004) which is by two orders of magnitude smaller than that for Mn$_{0.2}$TiSe$_2$.  The deformation of the large magnitude should increase the splitting of the lower unfilled Ti 3d$_{z^2}$-orbital from the rest of Ti 3d states, i.e. increase the binding energy of Ti d$_{z^2}$ orbital \cite{Bayliss:1984p8466,Gamble:1974p5931,Bullett:1978p6737,Huisman:1971p8447,Sugiura:1991p8389} and, consequently,  result in the partial filling of the electronic states in the vicinity of the Fermi level. 
 
As shown in Fig. \ref{fig:4}, Ti L$_{3,2}$ absorption spectrum calculated for Ti$^{4+}$ is in a good agreement with experimental data. In the calculated spectrum the main maxima \textbf{A} and \textbf{B } correspond to the mass centers of the e$_g$ and t$_{2g}$  orbitals, though the orbitals are strongly mixed. The splitting into antibonding t$_{2g}$  and  e$_g$ orbitals of Ti L$_3$ and Ti L$_2$ components of the multiplet in the calculated crystal field is equal to 1.2 eV (Fig. \ref{fig:4})  per contra to the case of TiSe$_2$ where the splitting is equal to 1.8 eV \cite{AShkvarin}. 
Calculated Ti L$_{3,2}$ absorption spectrum does not reproduce the peak \textbf{C} observed in the experimental spectrum. Since the L$_3$ absorption spectrum of titanium, similarly to the Ti 2p photoelectron spectrum,  is a superposition of weighted contributions from Ti1 and Ti2 atoms, the peak \textbf{C} is due to spectral contribution of Ti2 atoms. Calculated absorption spectrum of Mn$^{2+}$ shown in Fig. \ref{fig:4} reproduces much better both the shape and the energy position of main spectral maxima and the energy splitting of t$_{2g}$  and e$_g$ orbitals of Mn is absent. 

\subsection{ResPES}
\label{sec:ResPES}
Resonant photoemission in TM$_x$TiZ$_2$ can be considered within the framework of common photoemission theory of metals. Photoemission of $d$ electrons from 3d metals and their compounds is strongly enhanced if the energy of the incident photon is slightly larger than the binding energy of  selected core level upon excitation of $np$ electrons ($n$=2,3) to an unfilled 3d state, resulting in a resonant electron emission  \cite{JWAllenbook,Kay:1998p5777,Rubensson:1997p1742,Martensson:1997p5868}. This effect is interpreted as a result of a coherent process in which an $np$ electron in the initial state is exited to an empty $3d$ level forming an intermediate bound state ($np$, $3d$). Because of the autoionization process the intermediate bound state can transform into a final state identical to that resulting from the direct photoemission. A resonance of 3d electrons takes place upon the interference of the direct channel of photoemission ($np^63d^n+h\nu\rightarrow np^63d^{n-1}e_f$) with the successive decay of the exited state of the Coster-Kronig type ($np^63d^{n+1}+h\nu \rightarrow np^53d^{n+1} \rightarrow np^63d^{n-1}e_f$).  Resonant peaks in a spectrum of valence band are superimposed by intense peaks of the Auger L$_{2,3}$M$_{4,5}$M$_{4,5}$ line. 

\begin{figure*}
\begin{center}
\includegraphics[width=0.75\textwidth] {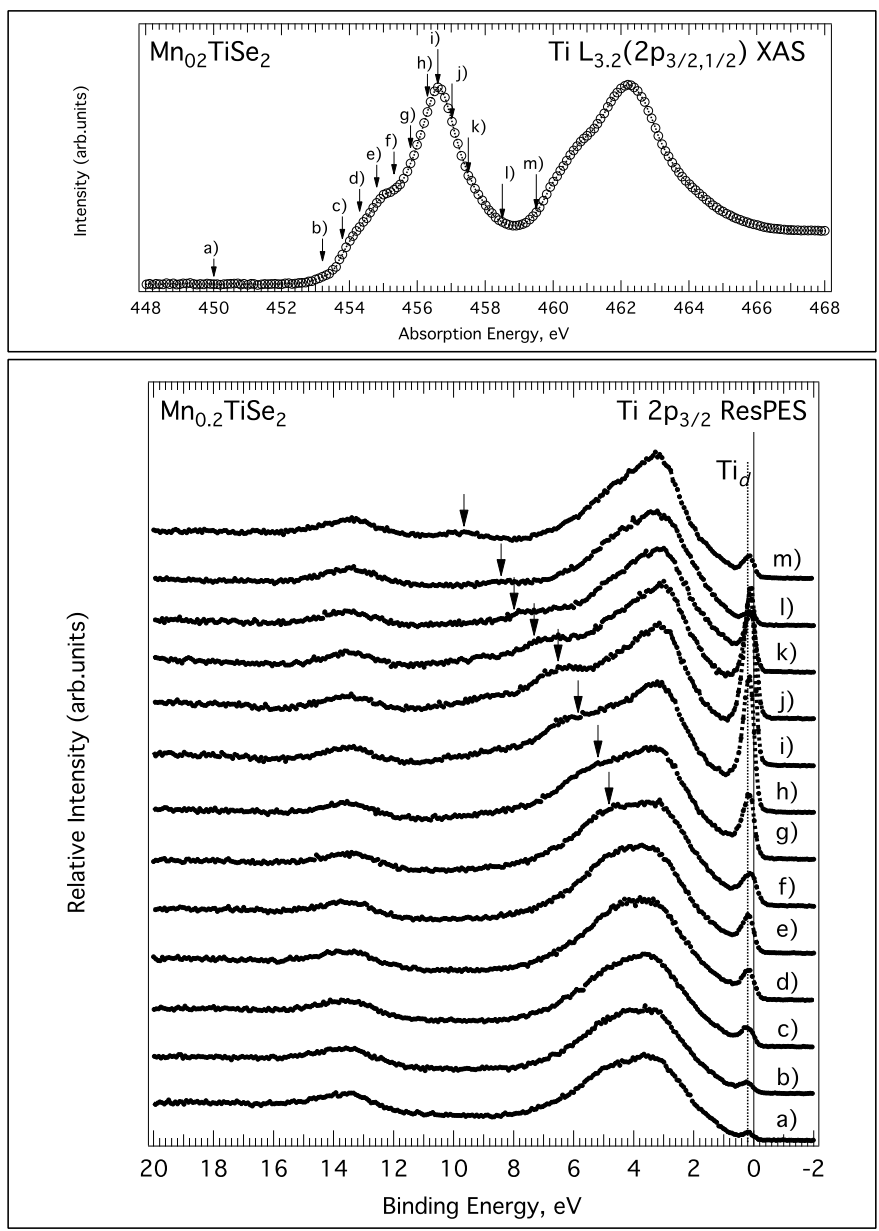}
\end{center}
\caption{ Ti 2p$_{3/2}$  valence band Resonant photoelectron spectra (ResPES) of Mn$_{0.2}$TiSe$_2$ are shown in the bottom frame of the figure. At the top frame Ti L$_{3,2}$ X-ray absorption spectrum is shown. Excitation energies are labelled by down-arrows in XAS and marked with letters from \textbf{a)} to \textbf{m)}. In the bottom frame, corresponding ResPES are labelled accordingly. Down-arrows in the bottom frame mark the positions of the Auger contribution. The peak with the binding energy of 13.5 eV corresponds to the energy position of Se 4s band.
}
\label{fig:5}       
\end{figure*}

\begin{figure*}
\begin{center}
\includegraphics[width=0.75\textwidth] {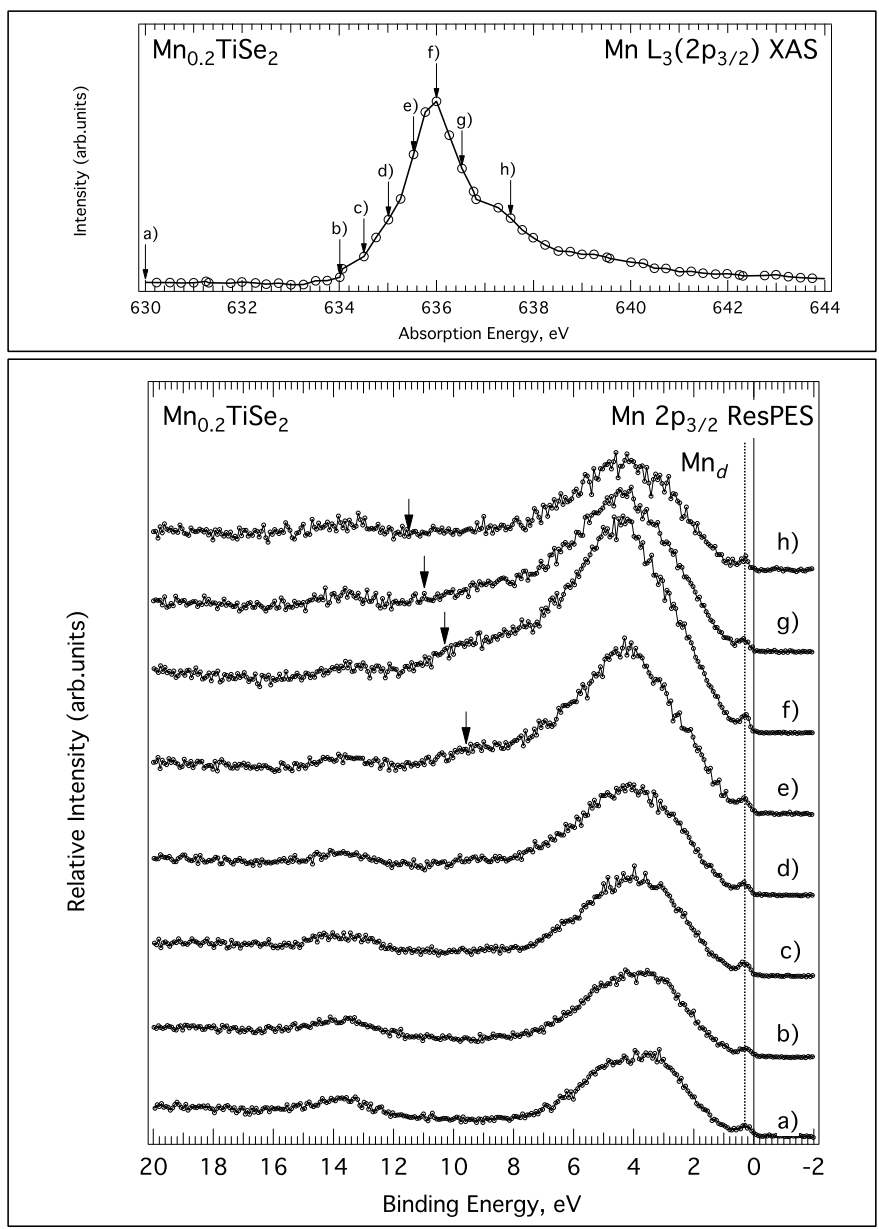}
\end{center}
\caption{Mn 2p$_{3/2}$  valence band ResPES and Mn L$_{3}$ X-ray absorption spectrum of Mn$_{0.2}$TiSe$_2$ are shown at the bottom and the top frames respectively. All designations are the same as in Fig. \ref{fig:5}.}
\label{fig:6}       
\end{figure*}

\begin{figure*}
\begin{center}
\includegraphics[width=0.85\textwidth] {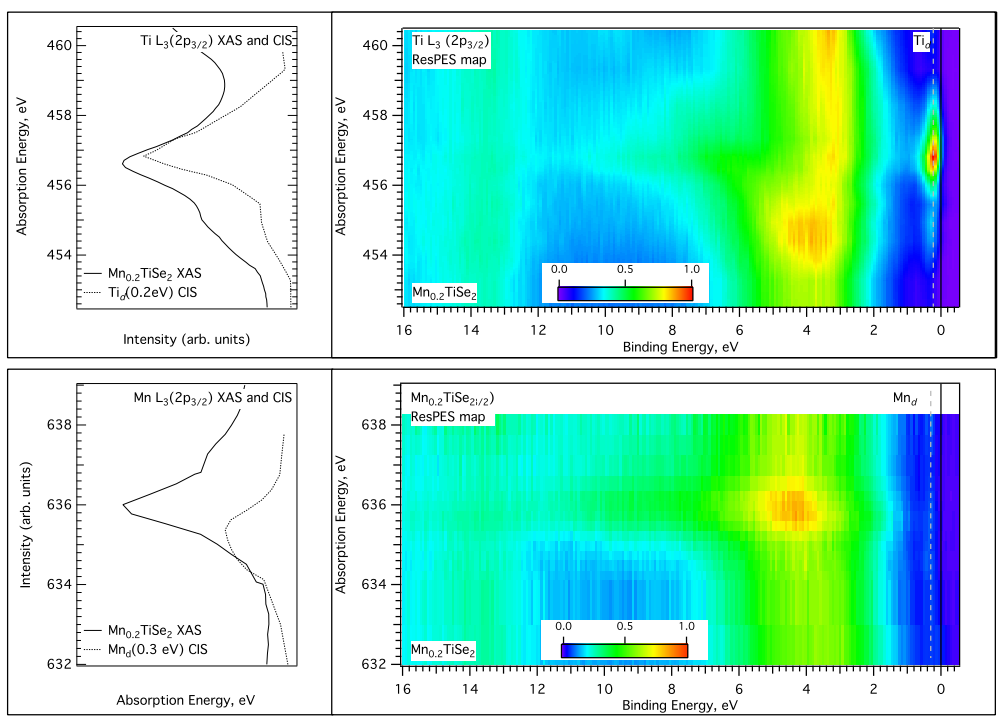}

\end{center}
\caption{ XAS, CIS (at the left side of the graph) and relative intensity maps of Ti 2p$_{3/2}$ and Mn 2p$_{3/2}$ ResPES interpolated out of data shown in Fig-s \ref{fig:5},\ref{fig:6} (right side of the graph).  At the top part of the figure the case of Ti excitation is presented. At the bottom part of the figure the case of Mn excitation is displayed. At the left side of the graph, X-ray absorption and constant intensity spectra of Ti$_d$ and Mn$_d$ intensity profiles are shown with solid and dotted lines respectively.  Position of Ti$_d$ and Mn$_d$ intensity profiles in the binding energy scale is marked with a dotted line in the righ-top and right-bottom frames respectively.}

\label{fig:9}       
\end{figure*}

ResPES of the valence band of Mn$_{0.2}$TiSe$_2$ taken in the range of excitation energies corresponding to those of Ti L$_3$ and Mn L$_3$ absorption edges (further labelled as Ti 2p$_{3/2}$ and Mn 2p$_{3/2}$ ResPES to indicate the element and the core level of excited electrons) are shown in Fig. \ref{fig:5} and  Fig. \ref{fig:6} respectively.  For both elements in cases of pre-resonant excitation a small broad peak in the range of 0.1-0.4 eV from the Fermi energy is observed, see spectra labelled by mark \textbf{a)} in Fig-s \ref{fig:5} and \ref{fig:6}. Following the increase of the excitation energy both standard Auger L$_3$VV spectrum of constant kinetic energy and resonance peaks in the vicinity of the Fermi level appear both in Ti 2p$_{3/2}$ and Mn 2p$_{3/2}$ ResPES.  In the normal mode of the Raman scattering (defined in the literature as Resonant Raman Auger Scattering - RRAS  \cite{JWAllenbook,Kay:1998p5777,Rubensson:1997p1742,Martensson:1997p5868}) the energy position of Auger signal in the binding energy scale is stable until the excitation energy is slightly less than the energy of maximum of TM L$_{3}$ X-ray absorption edge.  The situation is similar to that already studied in details for some 3d metals in Ref. \cite{Armen:1995p5870} where it was established that the crossover (transition from the RRAS to Auger mode) is different in energy for Cr, Fe, and Ni, i.e. is an element dependent. In our experiments the appearance of the crossover in resonant Ti 2p$_{3/2}$  ResPES is occurred  when the excitation energy reaches  the first spectral maximum of Ti L$_3$ absorption spectrum ( label \textbf{f)} in Fig. \ref{fig:5}). The Auger contribution is indicated by peaks shown by down-arrows in ResPES in Fig. \ref{fig:5}. Similar situation is observed in Mn 2p$_{3/2}$ ResPES, see in Fig. \ref{fig:6}. In both cases it is fortunate and important for further analysis that Auger peaks do not overlap with the peaks of valence band in the vicinity of the Fermi energy, since in that region a strong resonance at the binding energy of Ti$_d$ peak and a weak resonance at the energy of Mn$_d$ peak are observed in Ti 2p$_{3/2}$ and Mn 2p$_{3/2}$ ResPES respectively. A nature of the resonances in Ti 2p$_{3/2}$  and Mn 2p$_{3/2}$ ResPES can be easily determined from constant initial states (CIS) spectra which are the intensity profiles of Ti$_d$ and Mn$_d$ peaks taken from Fig-s \ref{fig:5}, \ref{fig:6} and displayed in X-ray absorption energy scales in Fig. \ref{fig:9}. As can be seen both Ti$_d$ and Mn$_d$ CIS include all peculiarities of L$_3$ XAS of Ti and Mn respectively indicating the presence of Ti 3d and Mn 3d states both located in the vicinity of the Fermi energy. 

\begin{figure*}
\begin{center}
 \includegraphics[width=0.9\textwidth]{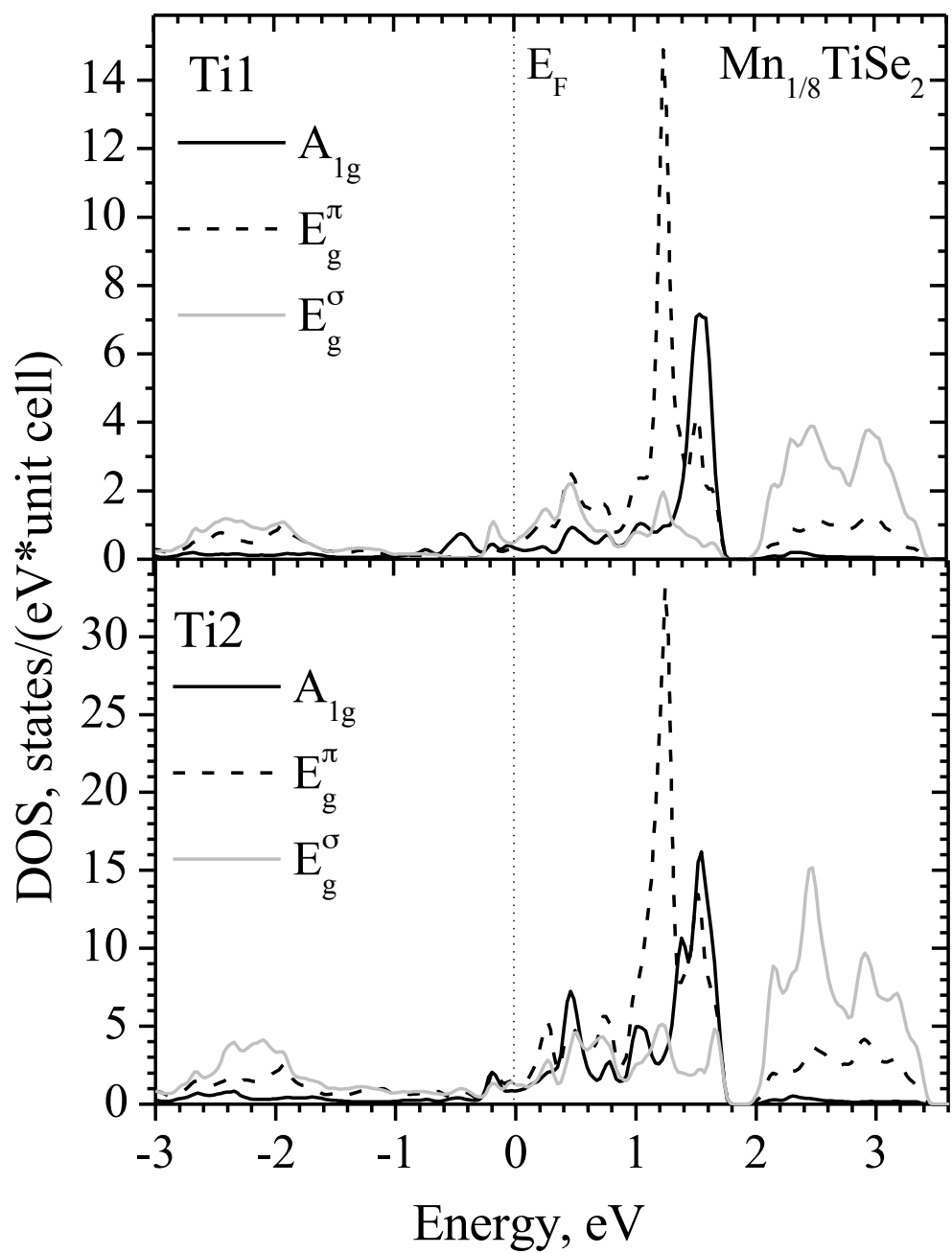}

\end{center}

\caption{
Local density of states Ti 3d in two nonequivalent positions Ti1 and Ti2. For Ti1 a relative increase of the A$_{1g} $ contribution to the states under the Fermi level in the energy range 0.2-0.3 eV is observed. Partial densities of states are given per formula unit.}
\label{fig:8}       
\end{figure*}

\begin{figure*}
\begin{center}
 \includegraphics[width=0.9\textwidth]{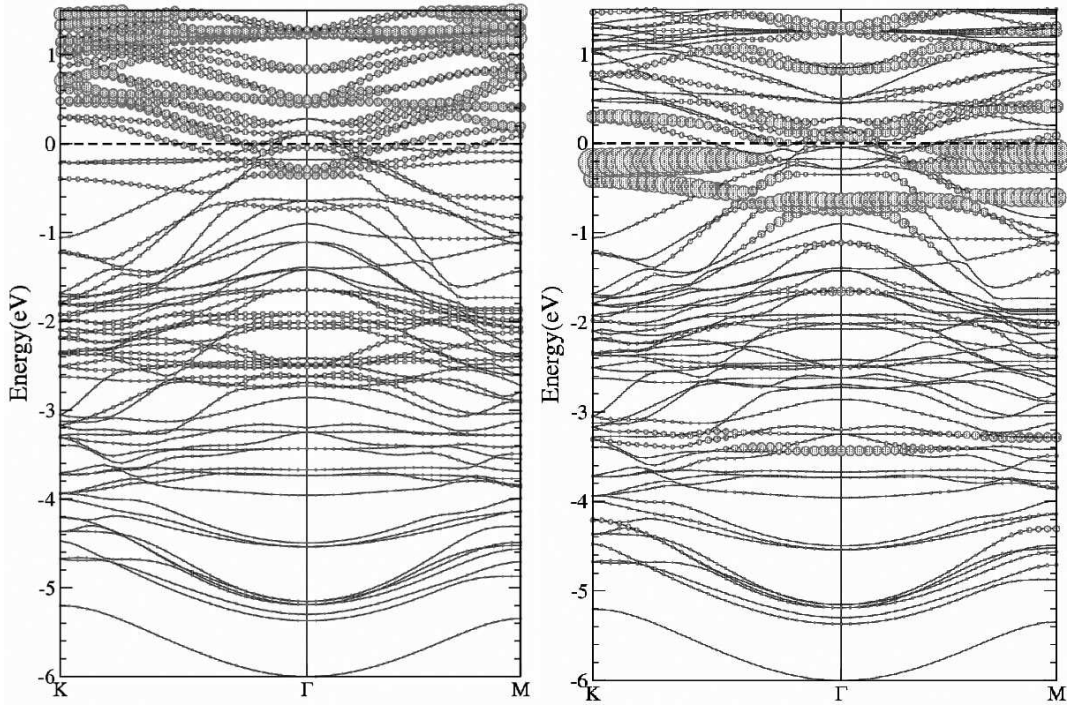}
\caption{Calculated dispersion curves for Mn$_{1/8}$TiSe$_2$ where, the circle diameter is proportional to the 3d zone filling by the titanium states (left panel) and manganese states (right panel). Under the Fermi level the 3d states of titanium  picked out in the vicinity of point $\Gamma$ mainly possess the mixed E$_g^\sigma$,  E$_g^\pi$ and A$_{1g}$ orbital symmetry ( energy 0.2 eV) . The Mn 3d states in directions  
$\mathsf{K}-\mathsf{\Gamma}-\mathsf{M}$
occupy the energy  range from the Fermi level up to 0.5 eV. They are  also possess  mixed  E$_g^\sigma$,  E$_g^\pi$  and A$_{1g}$  orbital symmetry. }
\label{fig:7}       
\end{center}
\end{figure*}

Since the influence of a hole in the valence band on a valence band spectrum is insignificant for case of 3d transition metals photoemission, it is reasonable to compare calculated DOS  and E(\textbf{k}) calculations shown in Fig-s \ref{fig:8} and \ref{fig:7}  with valence band photoemission spectra shown in Fig-s \ref{fig:5} and \ref{fig:6}.  A precise match between theory and experiment should not be expected due to following reasons. For  excitation energies in the ranges of 2p resonances of Ti and Mn the electron inelastic mean free path does not exceed 10 \AA \cite{Powell:1999p8582}. Taking into account that the magnitude of lattice constant in normal to basal plane direction equal to 6.15 \AA~for $x$=0.2 we estimate that the nearest to the crystal surface Van der Waals gap containing Mn atoms is located at a distance of  around 12 \AA. The distance is really close to the limits of electron inelastic mean free path. Besides, the crystal surface may lose a part of Mn atoms during the cleavage. Also correct computational data are available only for certain TM concentrations $x$ for which the modeling is correct, see Sec. \ref{sec:Experiment and calculations}. Nevertheless,  we have achieved a good match between the theory and the experiment. Therefore, the calculation results  are considered to be feasible for the following comparison of the calculated band structure with the experimental data. 

Because of the reasons mentioned above it is feasible to evaluate the origin of photoelectron resonance only for Ti 2p$_{3/2}$ ResPES in Mn$_{0.2}$TiSe$_2$. From the calculated local density of 3d states of titanium shown in Fig. \ref{fig:8}  it is clear that the electronic band represented by Ti$_d$ peak consists of two electronic bands both located near the Fermi level at slightly different binding energies.  The first band, which consists of completely hybridized orbitals, is found at the energy -0.2 eV whereas the second band, which posses symmetry A$_{1g}$,  is situated at the energy -0.4 eV. The filling of the lower band in the partial density of 3d states of Ti1 atom is remarkably enhanced as can be seen in Fig. \ref{fig:8}.  
Due to the energy resolution of experimental spectra these two contributions of Ti1 and Ti2 atoms in valence band ResPES are observed as one peak labelled as Ti$_d$ in Fig-s \ref{fig:5} and \ref{fig:9}, where the contribution of Ti2 atoms to the amplitude of Ti$_d$ peak should exceed the contribution of Ti1 atoms. To illustrate that the Ti$_d$ peak represents electronic states of different symmetry let us consider excitation dynamics of Ti 2p$_{3/2}$ ResPES shown in Fig. \ref{fig:5}. At the excitation energies from $\textbf{b)}$ to $\textbf{f)}$  in Ti L$_{2,3}$ XAS, primarily the t$_{2g}$  states (at octahedral symmetry) are excited whereas when increasing the excitation energy from  $\textbf{f)}$ to $\textbf{i)}$ the contribution of e$_g$ states is increased. From those considerations it is concluded that the Ti$_d$ peak represents the Ti band mixed of t$_{2g}$ and e$_g$ states.  

Amplitudes of Ti$_d$ and Mn$_d$ resonant peaks may be also influenced by experimental geometry. According to the theoretical calculations of E(\textbf{k}) shown in Fig.  \ref{fig:7} the Ti 3d states localized in the vicinity of $\mathsf{\Gamma}$ point prevail over the Mn 3d states localized primarily near the points K and M of the Brillouin zone at the energies just below the Fermi level. Although the value of photoionization cross-section of Ti 3d is approximately 3-4 times large then that of Mn 3d \cite{Yeh:1985p14258} at the energy of corresponding 2p threshold and the concentration of Mn is factor five smaller than that of Ti, the sample orientation may also influence the intensity of Ti$_d$ resonant peak in Ti 2p$_{3/2}$ ResPES. In the chosen experimental setup, in which the sample was oriented in a way that the \textit{c} axis of the crystal is parallel to the analyzer's axis, we believe that the experimental geometry creates favorable conditions for intensive electron emission from the vicinity of $\mathsf{\Gamma}$ point for the crystal of Mn$_{0.2}$TiSe$_2$. 
    
\section{Conclusion}

\label{Sec:Conclusion}
In Mn$_x$TiSe$_2$ compounds the chemical bonding between the host lattice and Mn atoms in the octahedral surrounding
is found to be of mixed ionic-covalent nature. These results are confirmed by the atomic multiplet calculations of Ti, Mn L$_3$ XAS and Mn 2p XPS.  According to  experimental data electrons are transferred from the manganese ions to the conduction band formed by Ti 3d states.

The Ti 3d and Mn 3d bands of mixed orbital symmetry have been observed at the binging energy just below the Fermi level in ResPES of the valence band. Performed E($\textbf{k}$) band structure calculations suggest that the Ti 3d states are localized in the vicinity of point $\mathsf{\Gamma}$ of the Brillouin zone of the crystal. The Mn 3d states are localized along the direction K-$\mathsf{\Gamma}$-M. 

The bonding situation for Mn$_x$TiSe$_2$ is found to be outstanding from that in non-Mn intercalated TM$_x$TiSe$_2$ due to the largest ionic contribution into chemical bonding between intercalated Mn and the host lattice.

\ack{
We acknowledge the financial support of the RFBR, grant N 09-03-00053.  We are thankful to F. de Groot for the opportunity to use his XAS code, Mikhail Korotin for useful recommendations, and Federica Bondino for introduction to the beamline and Michele Zacchigna for fruitful scientific discussion and careful corrections. 
}

\section*{References}
\bibliographystyle{iopart-num}
\bibliography{paper2011.07.14}

\newpage
\section*{Figures} 
%

\end{document}